\documentclass[conference]{IEEEtran}
\IEEEoverridecommandlockouts
\usepackage{cite}
\usepackage{amsmath,amssymb,amsfonts}
\usepackage{algorithmic}
\usepackage{algorithm}
\usepackage{graphicx}
\usepackage{caption}
\usepackage{subcaption}
\usepackage{textcomp}
\usepackage{xcolor}
\def\BibTeX{{\rm B\kern-.05em{\sc i\kern-.025em b}\kern-.08em
    T\kern-.1667em\lower.7ex\hbox{E}\kern-.125emX}}

\usepackage[normalem]{ulem}
\usepackage{xurl}
\usepackage[hidelinks]{hyperref}

\begin{document}

\title{STR-Agent: An LLM-Driven Agent for QoS-Aware Routing in LEO Satellite Networks
\thanks{\begin{tabular}[t]{@{}l@{}}
*Corresponding author.\\
Code is available at: 
\href{https://github.com/IntelliSensing/STR-Agent}
{\nolinkurl{https://github.com/IntelliSensing/STR-Agent}}.
\end{tabular}}
}

\author{
\IEEEauthorblockN{Bowen Lu, Mugen Peng, Yaohua Sun, Hongyu Wang, Kerui Guo, and Wenjia Xu\textsuperscript{*}}
\IEEEauthorblockA{
State Key Laboratory of Networking and Switching Technology\\
Beijing University of Posts and Telecommunications\\
Email: xuwenjia@bupt.edu.cn, bowenlu@bupt.edu.cn
}
}

\maketitle

\begin{abstract}
LEO satellite networks feature dynamic topologies, time-varying links, and diverse service requirements, which make conventional routing schemes difficult to support fine-grained quality-of-service (QoS) provisioning. Existing studies mainly optimize routing over network states with predefined objectives, but rarely address the practical challenge of translating unstructured natural-language service requests into adaptive routing decisions. To bridge this gap, we propose STR-Agent, an LLM-driven framework for QoS-aware routing in LEO satellite networks. The key innovation of STR-Agent lies in unifying intent perception, tool-based execution, experience accumulation, and reflection-based policy adaptation within a single agent architecture. Specifically, the Perception Module converts natural-language requests into structured routing semantics, while the Reflection Module dynamically adjusts the service-to-routing-policy mapping according to real-time congestion conditions and historical routing outcomes, rather than relying on a fixed routing objective. In addition, we develop a specialized perception model, and construct a domain-specific supervised fine-tuning dataset for LEO service understanding. Simulation results in a Walker–Delta constellation show that STR-Agent significantly outperforms conventional baselines: it reduces end-to-end delay by up to 60\% compared with DQ-Dijkstra, improves average intent-understanding accuracy from 45.4\% to 92.45\% after supervised fine-tuning, and the Reflection Module further reduces the delay by 120 ms at 600 Mbps. These results demonstrate the potential of LLM-driven agent architectures to enable service-aware and adaptive QoS routing in future LEO satellite networks.

\end{abstract}


\section{Introduction}

With the large-scale deployment of Low Earth Orbit (LEO) satellite constellations, LEO systems are becoming critical infrastructure for global communications \cite{b1}. Compared with Medium Earth Orbit (MEO) and Geostationary Earth Orbit (GEO) satellites, they provide superior deployment flexibility, enhanced spectrum efficiency, and reduced link loss \cite{b2}. Many LEO initiatives therefore aim to deploy tens of thousands of satellites and integrate them with terrestrial cellular networks for seamless global connectivity \cite{b3}. However, LEO networks are inherently dynamic, with high node mobility and rapidly varying inter-satellite and satellite-to-ground links \cite{b4}, fundamentally differing from relatively static terrestrial networks. As a result, terrestrial routing mechanisms often cannot adapt to topology evolution and link variation, leading to path mismatch, degraded forwarding efficiency, and unstable service quality \cite{b5}. Therefore, routing strategies with real-time network awareness and adaptive decision-making are essential.

Recent advances in artificial intelligence have shifted LEO routing from conventional static approaches to data-driven, learning-based methods \cite{b6}. The focus has moved from fixed rule-based path selection to learning network state, predicting link dynamics, and optimizing routing decisions \cite{b7}. Reinforcement learning-based routing algorithms, in particular, offer greater adaptability in dynamic environments \cite{b8}. However, existing learning-based routing methods in LEO networks still face two important limitations. First, their optimization objectives are typically predefined and fixed during training, making it difficult to flexibly balance latency, throughput, load balancing, and reliability across heterogeneous services \cite{b9}. As terrestrial and satellite networks become increasingly integrated, user applications exhibit growing diversity, and different services impose distinct QoS requirements \cite{b10}. Second, most methods decouple routing optimization from service understanding, assuming service type, priority, or QoS requirements are structured inputs, whereas user demands are often expressed in natural language. Existing approaches can optimize predefined objectives, but cannot interpret service intent, extract key constraints, or translate them into QoS-aware routing actions \cite{b11,b12}. This limitation is more pronounced in integrated terrestrial--satellite networks, where fine-grained QoS provisioning requires both path optimization and accurate service understanding. The key challenge is therefore to bridge natural-language service semantics and QoS-aware routing actions within a unified framework.

To address this issue, we formulate QoS-aware routing in LEO networks as a closed-loop intent-to-routing problem, where natural-language requests are translated into structured service semantics and then mapped to adaptive routing policies based on network congestion and historical routing outcomes. Based on this, we propose the LLM-driven Satellite Task-aware Routing Agent (STR-Agent), integrating Perception, Execution, Experience Buffer, and Reflection into a unified perceive–execute–reflect framework for service understanding, routing, experience summarization, and policy refinement. To improve domain understanding in LEO routing scenarios, we further construct a domain-specific dataset for service-request understanding and develop a specialized perception model.

The main contributions are summarized as follows:
\begin{itemize}
    \item We propose STR-Agent, a closed-loop intent-to-routing framework for LEO satellite networks, which transforms natural-language service requests into structured routing semantics and further into QoS-aware routing actions through coordinated perception, execution, experience buffering, and reflection.
    \item We propose an adaptive service-to-algorithm mapping mechanism that leverages both the current congestion profile and summarized routing experience to dynamically assign routing policies to different service types, rather than relying on a fixed routing objective.
    \item We construct a domain-specific supervised fine-tuning dataset for LEO service-request understanding, and validate through intent-understanding and routing experiments the contribution of the perception and reflection modules to QoS-aware routing performance.
\end{itemize}

\begin{figure}[t]
    \centering
    \includegraphics[width=0.98\columnwidth]{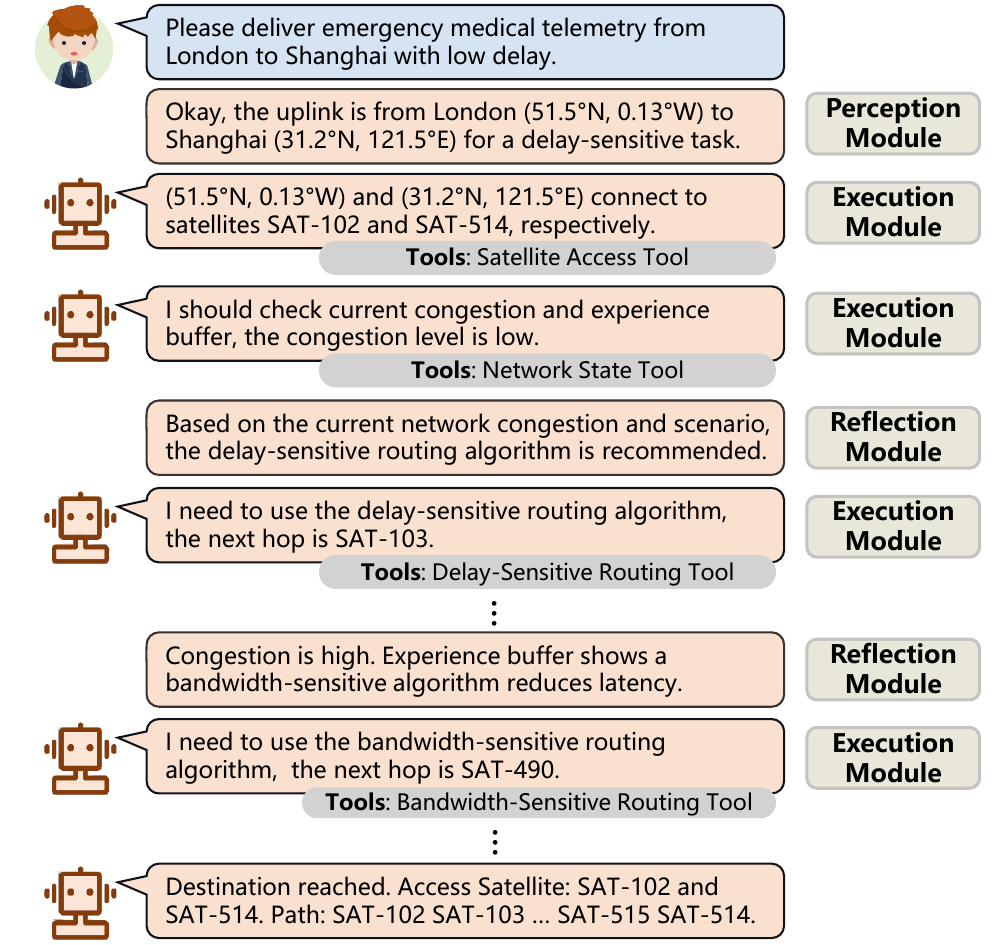}
    \caption{Workflow of STR-Agent for QoS-aware routing in LEO satellite networks.}
    \label{fig:teaser}
\end{figure}

Fig.~\ref{fig:teaser} illustrates the STR-Agent workflow. Given a natural-language service request, STR-Agent identifies the service intent, and performs multi-hop routing through tool execution and reflection-driven strategy adjustment.

\section{System Model}

We consider a Walker--Delta low Earth orbit (LEO) mega-constellation for global broadband services. The network is modeled as a time-varying graph with packet forwarding over inter-satellite links (ISLs). The constellation, link, delay, and QoS-oriented routing objective are defined as follows.

\subsection{Constellation Model}

The constellation is denoted by $(N_p,N_m,F)$, where $N_p$ is the number of orbital planes, $N_m$ is the number of satellites per plane, and $F$ is the phasing factor. The total number of satellites is $N=N_pN_m$. Each satellite operates at altitude $H$ with orbital radius $r=R_e+H$, where $R_e$ is the Earth radius.

At time $t$, the network is represented as $G(t)=(V,E(t))$, where $V$ and $E(t)$ denote the satellite set and active ISL set, respectively. Each satellite maintains four links, including two intra-orbit links within the same plane and two inter-orbit links to adjacent planes, forming a regular mesh topology.

\subsection{Link Model}

For any active link $(i,j)\in E(t)$, let $\mathbf{p}_i(t),\mathbf{p}_j(t)\in\mathbb{R}^3$ denote the positions of satellites $i$ and $j$. The inter-satellite distance is $d_{ij}(t)=\|\mathbf{p}_i(t)-\mathbf{p}_j(t)\|_2$. Assuming free-space propagation, the path loss is
\begin{equation}
L_{ij}(t)=\left(\frac{4\pi f_c d_{ij}(t)}{c}\right)^2,
\end{equation}
where $f_c$ is the carrier frequency and $c$ is the speed of light. The received power is $P_{r,ij}(t)=P_tG_tG_r/L_{ij}(t)$, where $P_t$, $G_t$, and $G_r$ denote the transmit power and the transmit and receive antenna gains, respectively.

Since practical inter-satellite links employ fixed modulation and coding schemes, we use an $E_b/N_0$-based model to approximate the achievable transmission rate~\cite{b13}:
\begin{equation}
R_{ij}(t)=\frac{P_{r,ij}(t)}{(E_b/N_0)\,k_B T_s},
\end{equation}
where $E_b/N_0$ is the required energy-per-bit to noise spectral density ratio, $k_B$ is the Boltzmann constant, and $T_s$ is the system noise temperature.

\subsection{Delay Model}

For a packet of size $L_p$ transmitted over link $(i,j)$ at time $t$, the propagation, transmission, and queuing delays are $\tau^{\mathrm{prop}}_{ij}(t)=d_{ij}(t)/c$, $\tau^{\mathrm{tx}}_{ij}(t)=L_p/R_{ij}(t)$, and $\tau^{\mathrm{que}}_{ij}(t)=L_pQ_{ij}(t)/R_{ij}(t)$, respectively, where $Q_{ij}(t)$ denotes the queue length at satellite $i$ for packets forwarded to satellite $j$. The total one-hop delay is
\begin{equation}
\tau_{ij}(t)=\tau^{\mathrm{prop}}_{ij}(t)+\tau^{\mathrm{tx}}_{ij}(t)+\tau^{\mathrm{que}}_{ij}(t).
\end{equation}

\begin{figure*}[t]
\centering
\includegraphics[width=160mm]{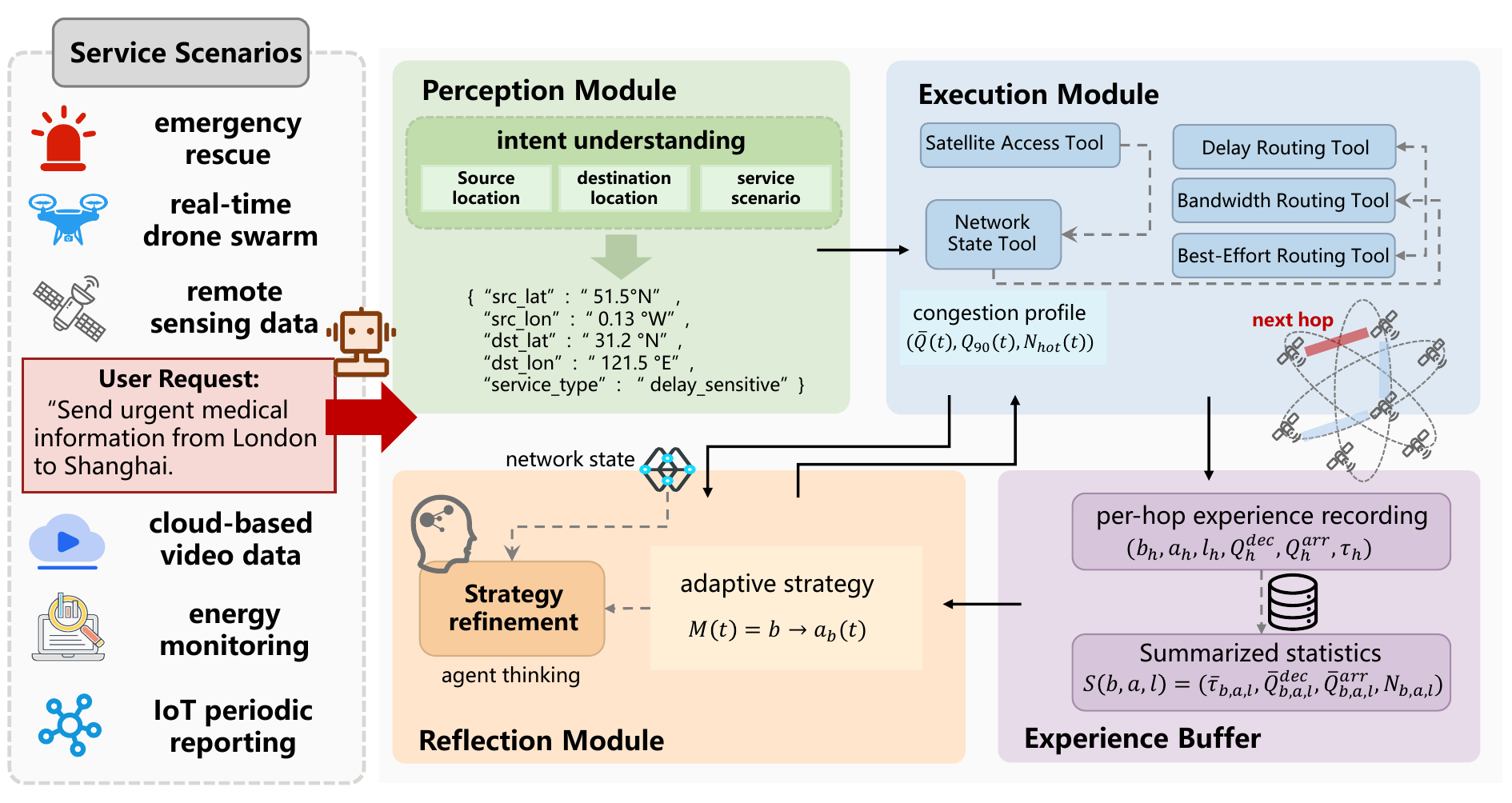}
\caption{Architecture of the proposed Satellite Task-aware Routing Agent (STR-Agent).}
\label{fig:framework}
\end{figure*}

\subsection{QoS-Oriented Routing Objective}

Based on the above model, QoS-aware routing is performed on the time-varying graph $G(t)$. For a service request $r_k$, the perception module extracts structured semantics and identifies the service type $b_k\in\{\mathcal{B}_{\mathrm{delay}},\mathcal{B}_{\mathrm{bw}},\mathcal{B}_{\mathrm{hop}}\}$. The routing module then outputs an end-to-end path
\begin{equation}
\mathcal{P}_k(t)=\{v_0,v_1,\ldots,v_m\}.
\end{equation}

Instead of optimizing a fixed global metric, STR-Agent adopts a service-dependent objective:
\begin{equation}
\mathcal{P}_k^\star(t)=\arg\min_{\mathcal{P}\in\Pi_k(t)} J(\mathcal{P}\mid b_k,G(t)),
\end{equation}
where $\Pi_k(t)$ is the feasible path set for request $r_k$, and $J(\mathcal{P}\mid b_k,G(t))$ is the service-adaptive path cost.

For delay-sensitive services, the objective is to minimize cumulative latency, i.e., $J_{\mathrm{delay}}(\mathcal{P})=\sum_{(i,j)\in\mathcal{P}}\tau_{ij}(t)$. For bandwidth-sensitive services, the objective is to avoid congested or highly imbalanced regions, which can be written as $J_{\mathrm{bw}}(\mathcal{P})=\sum_{(i,j)\in\mathcal{P}}\sigma(\{Q_n(t):n\in\mathcal{N}(i)\})$, where $\mathcal{N}(i)$ denotes the neighbor set of node $i$ and $\sigma(\cdot)$ denotes the standard deviation of queue lengths. For best-effort services, the objective is to reduce routing complexity and forwarding overhead, approximated by the hop count $J_{\mathrm{hop}}(\mathcal{P})=|\mathcal{P}|-1$.

This formulation provides the modeling basis for the routing tools in Section~\ref{sec:framework}: the routing policy is determined not by a fixed objective, but by a service-adaptive cost selected according to the inferred service type.

\section{Satellite Task-aware Routing Agent}
\label{sec:framework}
We propose the Satellite Task-aware Routing Agent (STR-Agent), an LLM-driven framework for QoS-aware routing in LEO satellite networks. As shown in Fig.~\ref{fig:framework}, STR-Agent consists of four modules: Perception for intent parsing, Execution for tool-grounded route construction, Experience Buffer for outcome statistics, and Reflection for adaptive policy selection. Together, they form a closed loop from natural-language requests to adaptive multi-hop forwarding.

\subsection{Perception Module}

Given natural-language request $r_k$, the Perception Module outputs
\begin{equation}
P(r_k)=\left(\phi_k^{\mathrm{src}},\lambda_k^{\mathrm{src}},\phi_k^{\mathrm{dst}},\lambda_k^{\mathrm{dst}},b_k\right),
\end{equation}
where $(\phi,\lambda)$ denote latitude and longitude, and $b_k$ denotes the inferred service type. This task contains two coupled subtasks: geographic entity extraction and service-type classification.

The service type $b_k$ is drawn from $\{\mathcal{B}_{\mathrm{delay}},\mathcal{B}_{\mathrm{bw}},\mathcal{B}_{\mathrm{hop}}\}$, corresponding to delay-sensitive, bandwidth-sensitive, and best-effort services, respectively. A typical output is
\begin{verbatim}
{"src_lat":"51.5°N","src_lon":"0.13°W",
 "dst_lat":"31.2°N","dst_lon":"121.5°E",
 "service_type":"delay_sensitive"}.
\end{verbatim}

This structured representation provides the geographic endpoints and service semantics required by downstream routing. Errors in service-type prediction may cause the Execution Module to invoke an unsuitable routing policy; therefore, intent understanding accuracy directly affects downstream QoS.

\subsection{Execution Module}

Given $P(r_k)$, the Execution Module first maps the source and destination coordinates to an access-satellite pair, and then performs routing through external tool calls. STR-Agent adopts hop-by-hop forwarding: at each step, it queries the current congestion profile, obtains the service-to-algorithm mapping from the Reflection Module, selects the routing tool, and computes the next hop. This enables online adaptation to time-varying congestion.

The Execution Module uses five tools: 
\begin{itemize}
  \item \textbf{Satellite Access Tool}, which maps ground coordinates $(\phi,\lambda)$ to the visible access satellite with the highest elevation angle;
  \item \textbf{Network State Tool}, which returns the current congestion profile
  \begin{equation}
  \mathcal{C}(t)=\bigl(\bar{Q}(t),Q_{90}(t),N_{\mathrm{hot}}(t)\bigr),
  \label{eq:congestion_profile}
  \end{equation}
  where $\bar{Q}(t)$ is the mean queue length, $Q_{90}(t)$ is the 90th-percentile queue length, and $N_{\mathrm{hot}}(t)$ is the number of satellites whose queue length exceeds a preset threshold;
  \item \textbf{Delay-Sensitive Routing Tool} $B_{\mathrm{delsy}}$;
  \item \textbf{Bandwidth-Sensitive Routing Tool} $B_{\mathrm{bw}}$;
  \item \textbf{Best-Effort Routing Tool} $B_{\mathrm{hop}}$.
\end{itemize}

All three routing algorithms are based on the Dijkstra Shortest Path (DSP) method, with edge weights varying according to service type. The weights are defined as follows.
For delay-sensitive traffic, the edge weight is
\begin{equation}
w_{ij}^{\mathrm{delay}}=\frac{d_{ij}}{c}+\frac{L_p}{R_{ij}},
\end{equation}
which favors low propagation and transmission delay.

For bandwidth-sensitive traffic, the edge weight is
\begin{equation}
w_{ij}^{\mathrm{bw}}=\sigma\!\left(\{Q_n:n\in\mathcal{N}(i)\}\right),
\end{equation}
where $\mathcal{N}(i)$ denotes the neighbor set of node $i$, $Q_n$ is the queue length of neighbor $n$, and $\sigma(\cdot)$ denotes the standard deviation. This metric encourages routing through regions with more balanced queue states and hence alleviates queue buildup for bandwidth-intensive traffic.

For best-effort traffic, the edge weight is
\begin{equation}
w_{ij}^{\mathrm{hop}}=1,
\end{equation}
which reduces to minimum-hop routing.

\subsection{Experience Buffer}
The Experience Buffer stores and summarizes hop-level routing outcomes. After each hop, STR-Agent records
\begin{equation}
e_h=\left(b_h,a_h,\ell_h,Q_h^{\mathrm{dec}},Q_h^{\mathrm{arr}},\tau_h\right),
\end{equation}
where $b_h$ is the service type, $a_h\in\{\mathcal{B}_{\mathrm{delay}},\mathcal{B}_{\mathrm{bw}},\mathcal{B}_{\mathrm{hop}}\}$ is the selected routing tool, $\ell_h$ is the congestion level at decision time, $Q_h^{\mathrm{dec}}$ and $Q_h^{\mathrm{arr}}$ are the queue lengths observed at forwarding and arrival, and $\tau_h$ is the realized one-hop delay.

To preserve congestion dependence, these records are aggregated as
\begin{equation}
S(b,a,\ell)=\left(\bar{\tau}_{b,a,\ell},\bar{Q}^{\mathrm{dec}}_{b,a,\ell},\bar{Q}^{\mathrm{arr}}_{b,a,\ell},N_{b,a,\ell}\right),
\end{equation}
where $\bar{\tau}_{b,a,\ell}$ denotes the mean delay, $\bar{Q}^{\mathrm{dec}}_{b,a,\ell}$ and $\bar{Q}^{\mathrm{arr}}_{b,a,\ell}$ denote the average queue lengths, and $N_{b,a,\ell}$ is the sample count. Each hop contributes one record, and the summary is updated in a rolling manner after each request or time window so that recent observations dominate policy adjustment.

\subsection{Reflection Module}
The Reflection Module maps the current network state and accumulated experience to a service-to-algorithm mapping
\begin{equation}
M(t): b \mapsto a_b(t),
\end{equation}
where $a_b(t)$ is the routing tool for service type $b$.

Under normal conditions, each service type uses its default routing tool. The Reflection Module regards the network as severely congested when the average satellite queue length is at least 5 or at least 50 satellites have queues longer than 5. It then uses the historical hop-level records in the latest reflection window: if both delay-sensitive and bandwidth-sensitive records exist and the bandwidth-sensitive records have a lower average per-hop delay, delay-sensitive traffic is temporarily switched to \(\mathcal{B}_{\mathrm{bw}}\); otherwise, the default delay-sensitive tool is kept. Bandwidth-sensitive and best-effort traffic continue to use \(\mathcal{B}_{\mathrm{bw}}\) and \(\mathcal{B}_{\mathrm{hop}}\), respectively.

\section{Model Fine-Tuning}

To enhance the Perception Module of STR-Agent, we fine-tune a domain-specific model for mapping natural-language service requests to structured routing semantics, with emphasis on geographic entity extraction and service-type recognition.

\subsection{Dataset Construction}
\label{subsec:dataset_construction}
We construct a supervised fine-tuning dataset tailored to LEO satellite networks. Satellite communication services are divided into three QoS-oriented categories, namely delay-sensitive, bandwidth-sensitive, and best-effort. For each category, 14 representative scenarios are selected with reference to 3GPP TR 22.822~\cite{b14}, covering typical LEO applications such as emergency communications, data backhaul, and large-scale IoT services. The geographic space is instantiated by 49 global cities, from which source--destination pairs are sampled to form realistic service requests.

All samples are generated by Qwen2.5-7B from scenario descriptions and city pairs in a structured \texttt{\{instruction, input, output\}} format. The \texttt{input} is a natural-language request containing the upstream and downstream locations and scenario context, while the \texttt{output} gives the corresponding source and destination coordinates and service type. The training set contains 30{,}000 samples, with 10{,}000 per class, and the test set contains 3{,}000 samples, with 1{,}000 per class.

\subsection{Supervised Fine-Tuning}
\label{subsec:sft}
We fine-tune Qwen2.5-7B using LoRA to adapt the model for mapping natural-language service requests to structured routing semantics. This enables the Perception Module to accurately interpret diverse service intents and provide reliable inputs to the Execution and Reflection Modules, enabling precise, routing-oriented decision-making.

\begin{figure*}[t]
\centering

\begin{subfigure}[t]{0.46\textwidth}
    \makebox[\linewidth][r]{%
        \begin{minipage}{0.76\linewidth}
            \centering
            \includegraphics[width=\linewidth]{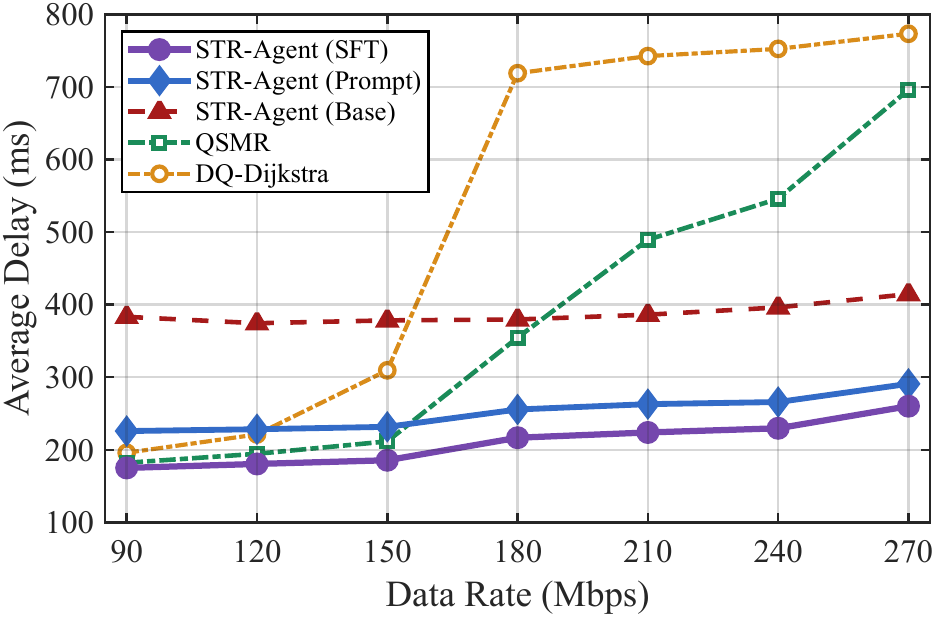}
            \captionsetup{justification=centering}
            \caption{Average end-to-end delay}
            \label{fig:end-to-end delay}
        \end{minipage}
    }
\end{subfigure}\hspace{0.02\textwidth}%
\begin{subfigure}[t]{0.46\textwidth}
    \makebox[\linewidth][l]{%
        \begin{minipage}{0.76\linewidth}
            \centering
            \includegraphics[width=\linewidth]{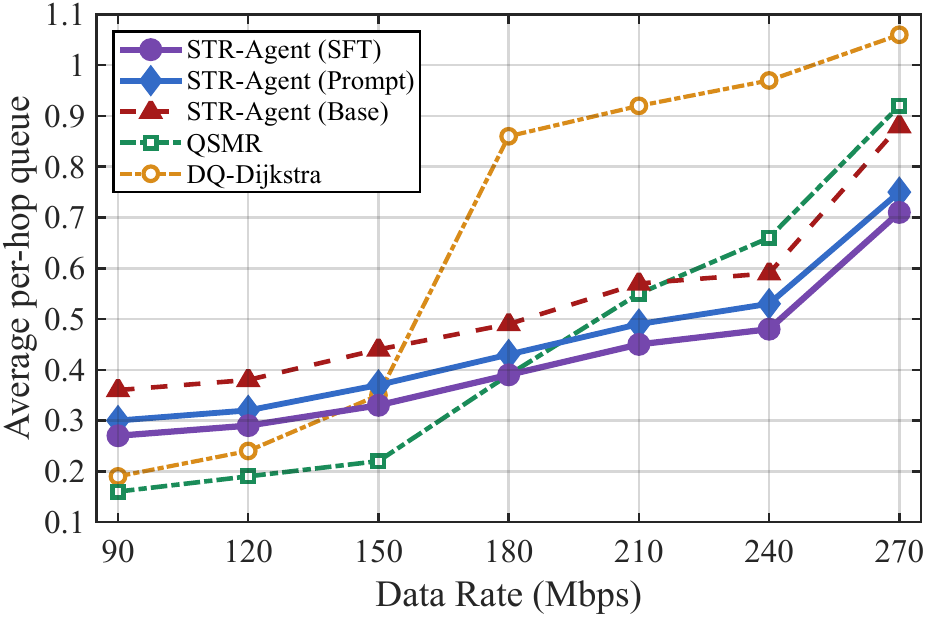}
            \captionsetup{justification=centering}
            \caption{Average per-hop queue length}
            \label{fig:per-hop queue}
        \end{minipage}
    }
\end{subfigure}

\vspace{0.012\textheight}

\begin{subfigure}[t]{0.46\textwidth}
    \makebox[\linewidth][r]{%
        \begin{minipage}{0.76\linewidth}
            \centering
            \includegraphics[width=\linewidth]{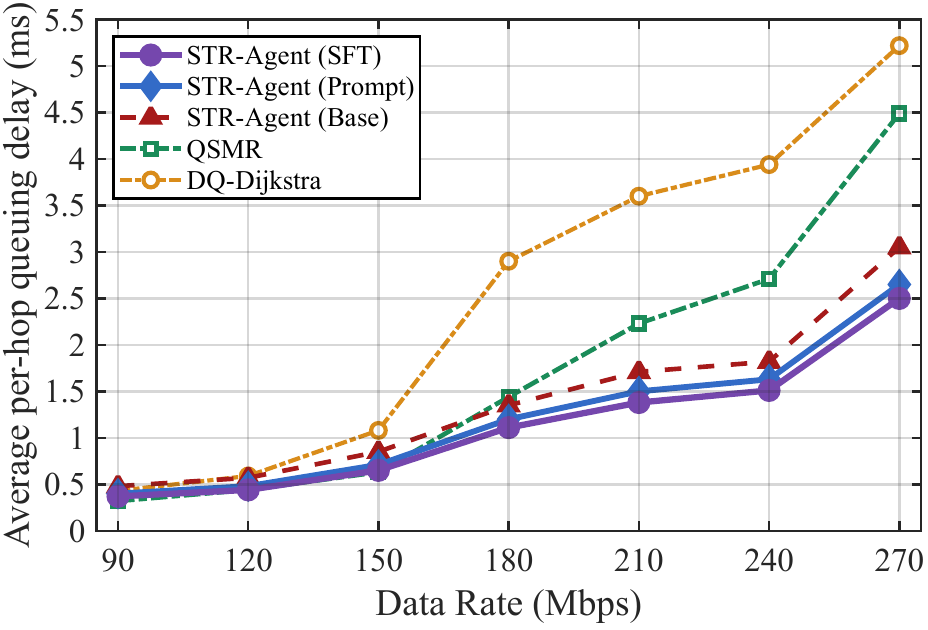}
            \captionsetup{justification=centering}
            \caption{Average per-hop queuing delay}
            \label{fig:per-hop queuing delay}
        \end{minipage}
    }
\end{subfigure}\hspace{0.02\textwidth}%
\begin{subfigure}[t]{0.46\textwidth}
    \makebox[\linewidth][l]{%
        \begin{minipage}{0.76\linewidth}
            \centering
            \includegraphics[width=\linewidth]{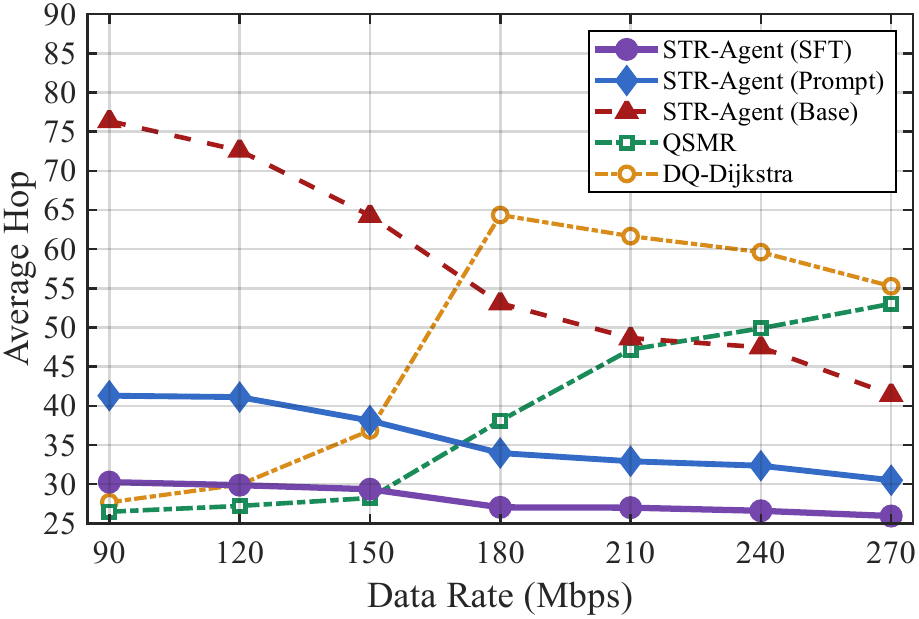}
            \captionsetup{justification=centering}
            \caption{Average hop count}
            \label{fig:hop count}
        \end{minipage}
    }
\end{subfigure}

\caption{(a) Average end-to-end delay, (b) average per-hop queue length, (c) average per-hop queuing delay, (d) average hop count under different data rates.}
\label{fig:p1}
\end{figure*}

\section{Experiments}

In this section, we evaluate STR-Agent through simulations in a Walker--Delta LEO constellation.

\subsection{Simulation Setup}

The constellation configuration is $(N_p,N_m,F)=(72,22,39)$, corresponding to 1584 satellites at an altitude of 550 km. The carrier frequency is 23.28 GHz, the channel bandwidth is 2 GHz, the transmit EIRP is 35 dBW, the receiver quality factor is $G_r/T_s=6.8$ dB/K, and the modulation scheme is 8-PSK with a target BER of $10^{-7}$. The LoRA rank is 8, the scaling factor is 16, the dropout rate is 0.05, the learning rate is $2\times10^{-4}$, the batch size is 64, and the number of training epochs is 3.

Service requests follow the dataset construction in Section~\ref{subsec:dataset_construction}. Performance is evaluated separately for different service types to reflect QoS-differentiated routing behavior. During routing, STR-Agent performs hop-by-hop forwarding: at each step, it queries the current congestion profile and selects the routing tool accordingly.

\begin{table}[h]
\caption{Intent Understanding Accuracy (\%)}
\label{tab:intent_accuracy}
\centering
\footnotesize
\renewcommand{\arraystretch}{1.1}
\setlength{\tabcolsep}{3.5pt}
\begin{tabular}{lcccc}
\hline
\textbf{Model} & \textbf{Delay-Sens.} & \textbf{Bandwidth-Sens.} & \textbf{Best-Effort} & \textbf{Avg.} \\
\hline
Base Model & 78.02 & 49.9 & 8.27 & 45.4 \\
Prompt-Based Model & 82.41 & 86.49 & 73.32 & 80.74 \\
SFT-Based Model & 97.7 & 95.45 & 84.2 & 92.45 \\
\hline
\end{tabular}
\end{table}

\subsection{Intent Understanding Evaluation}
\label{subsec:intent_eval}
We first evaluate the perception module on natural-language service requests. The metric is intent-understanding accuracy, defined as the proportion of requests for which the source, destination, and service type are all correctly identified. The test set is generated following Section~\ref{subsec:dataset_construction}.

We compare three models: the \textit{Base Model}, namely the original Qwen2.5-7B; the \textit{Prompt-Based Model}, which uses task-specific prompts; and the \textit{SFT-Based Model}, which is fine-tuned as described in Section~\ref{subsec:sft}. As shown in Table~\ref{tab:intent_accuracy}, the Base Model achieves only 45.4\% average accuracy, indicating limited domain-specific parsing ability. Prompting improves the average accuracy to 80.74\%, and supervised fine-tuning further raises it to 92.45\%.

Best-effort requests are consistently harder than the other two categories. In the Base Model, their accuracy is only 8.27\%, far below the delay-sensitive and bandwidth-sensitive cases, mainly because their weaker semantic cues make them more easily confused with other scenarios.

These results directly affect downstream routing, since misclassified service types lead to unsuitable routing-tool selection, explaining the routing gap among the Base, Prompt-Based, and SFT-Based models.

\subsection{Routing Performance Evaluation}  
\label{subsec:routing_eval}  
We evaluate routing performance under varying traffic conditions, with traffic arrivals modeled as a Poisson process. The data rate ranges from 90 to 270~Mbps, covering low- to high-load conditions, and each simulation lasts $T=10{,}000$~ms. Performance is evaluated separately for different service types to reflect QoS-differentiated routing behavior.

We compare the following methods:
\begin{itemize}
    \item \textbf{DQ-Dijkstra}: DSP with edge weights combining delay and queue length, balancing latency and congestion.
    \item \textbf{QSMR~\cite{b15}}: A multi-routing algorithm that jointly considers propagation delay and residual link capacity.
    \item \textbf{STR-Agent (Base)}: Base-model STR-Agent.
    \item \textbf{STR-Agent (Prompt)}: Prompt-based STR-Agent.
    \item \textbf{STR-Agent (SFT)}: SFT-based STR-Agent.
\end{itemize}

We use four metrics corresponding to different service types: average end-to-end delay for delay-sensitive traffic, average per-hop queue length and per-hop queuing delay for bandwidth-intensive traffic, which reflect the load-balancing capability of the method, and average hop count for best-effort traffic, which evaluates routing simplicity and efficiency.

Fig.~\ref{fig:end-to-end delay} shows that \textit{STR-Agent (SFT)} consistently achieves the lowest delay over the entire load range. At 270 Mbps, it reduces the delay from 696.01 ms with \textit{QSMR} to 260.11 ms, while \textit{STR-Agent (Prompt)} also surpasses \textit{DQ-Dijkstra} and \textit{QSMR} from 180 Mbps onward. Fig.~\ref{fig:per-hop queue} and Fig.~\ref{fig:per-hop queuing delay} show that, although \textit{DQ-Dijkstra} and \textit{QSMR} perform better at low load, \textit{STR-Agent (SFT)} outperforms the baselines as the load increases and, at 270 Mbps, reduces the average per-hop queue length from 1.06 to 0.71 and the per-hop queuing delay from 5.22 ms to 2.50 ms compared with \textit{DQ-Dijkstra}, indicating stronger load balancing under congested conditions. Fig.~\ref{fig:hop count} shows that \textit{STR-Agent (SFT)} achieves the smallest hop count at medium and high load, reducing it from 53.03 with \textit{QSMR} to 25.94 at 270 Mbps. Overall, among the three agent variants, \textit{STR-Agent (SFT)} performs best, \textit{STR-Agent (Prompt)} ranks second, and \textit{STR-Agent (Base)} yields the weakest performance, indicating that more accurate service understanding leads to better routing-policy selection and reduced routing-policy mismatch, especially under moderate and high traffic load.

\subsection{Reflection Module Evaluation}
\label{subsec:ablation}

\begin{figure}[h]
\centering
\includegraphics[width=60mm]{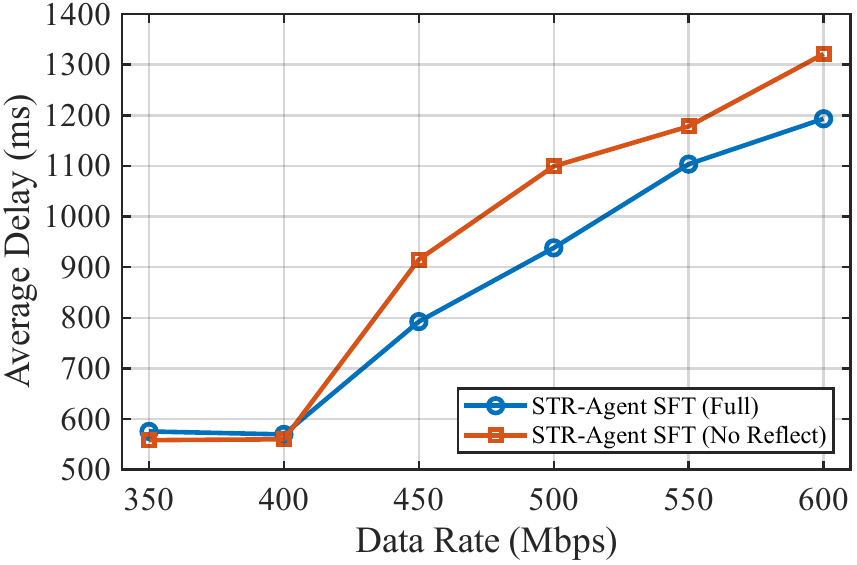}
\caption{Ablation study of Reflection Module for average delay in Delay-Sensitive traffic.}
\label{fig:reflection module}
\end{figure}

We further evaluate the Reflection Module under high-load conditions by comparing \textit{STR-Agent SFT (Full)} and \textit{STR-Agent SFT (No Reflect)} with the offered load, modeled as a Poisson process, ranging from 350 to 600 Mbps. In the latter setting, each service type always uses its default routing tool.

Fig.~\ref{fig:reflection module} shows that the Reflection Module improves delay mainly under medium- to high-load conditions. At 350--400~Mbps, the two configurations exhibit nearly identical delays, indicating limited benefit under light congestion. As the load increases, the gap becomes more pronounced, and at 600 Mbps the average delay decreases from 1320 ms to 1200 ms. Overall, \textit{STR-Agent SFT (Full)} consistently achieves lower delay, with gains increasing as congestion intensifies.

The Reflection Module enables STR-Agent to adapt routing based on historical experience and real-time congestion, avoiding overloaded paths, which confirms its importance for delay-sensitive tasks under high-load conditions.

\section{Conclusion}
This work shows that QoS-aware routing in LEO satellite networks can be effectively addressed by STR-Agent, a closed-loop Satellite Task-aware Routing Agent that unifies perception, execution, experience buffering, and reflection in a QoS-aware control framework. Rather than relying on a fixed routing objective, STR-Agent adapts routing decisions to heterogeneous service demands and changing congestion conditions. Experimental results show significant gains in delay and load-balancing performance over conventional baselines. Overall, STR-Agent provides a practical direction for adaptive and service-aware routing in future LEO satellite networks. Future work will examine human-labeled requests, learning-based baselines, and runtime overhead.

\section{Acknowledgement}
This work has been funded by the National Natural Science Foundation of China under Grant 62301063.

\end{document}